\documentclass[12pt,draftcls,onecolumn] {IEEEtran}
\usepackage[T1]{fontenc}
\ifCLASSINFOpdf
\else
\fi
\usepackage{amsmath}
\usepackage{cite}
\usepackage{amssymb}
\usepackage{subfigure}
\usepackage{graphicx}
\usepackage[cmintegrals]{newtxmath}
\begin{document}

\title{The Optimal Pilot Power Allocation Strategy for Multi-IRS Assisted Communication Systems}
\author{Jiancheng~An,~
        Chao~Xu,~\IEEEmembership{Senior~Member,~IEEE,}~
    Lu~Gan,~\\
    Chau~Yuen,~\IEEEmembership{Fellow,~IEEE,}~
        and~Lajos~Hanzo,~\IEEEmembership{Fellow,~IEEE}
\thanks{L. Hanzo would like to acknowledge the financial
support of the Engineering and Physical Sciences Research Council projects
EP/N004558/1, EP/P034284/1, EP/P034284/1, EP/P003990/1 (COALESCE),
of the Royal Society’s Global Challenges Research Fund Grant as well as of
the European Research Council’s Advanced Fellow Grant QuantCom.}
\thanks{J. An and L. Gan are with the School of Information and Communication Engineering, University of Electronic Science and Technology of China, Chengdu, Sichuan 611731, China. (e-mail: jiancheng$\_$an@163.com; ganlu@uestc.edu.cn).}
\thanks{C. Xu and L. Hanzo are with the School of Electronics and Computer Science, University of Southampton, Southampton, SO17 1BJ, U.K. (e-mail: cx1g08@soton.ac.uk; lh@ecs.soton.ac.uk).}
\thanks{C. Yuen is with the Engineering Product Development (EPD) Pillar, Singapore University of Technology and Design, Singapore 487372 (e-mail: yuenchau@sutd.edu.sg).}}

\maketitle

\begin{abstract}
Intelligent reflecting surface (IRS) is a promising technology that enables the precise control of the electromagnetic environment in future wireless communication networks. To leverage the IRS effectively, the acquisition of channel state information (CSI) is crucial in IRS-assisted communication systems, which, however, is challenging. In this paper, we propose the optimal pilot power allocation strategy for the channel estimation of IRS-assisted communication systems, which is capable of further improving the achievable rate performance with imperfect CSI. More specifically, \emph{first of all}, we introduce a multi-IRS assisted communication system in the face of practical channel estimation errors. Furthermore, the ergodic capacity with imperfect CSI is derived in an explicit closed-form expression under the single-input single-output (SISO) consideration, which is dependent on the specific pilot power for estimating different reflection channels. \emph{Secondly}, we formulate the optimization problem of maximizing the ergodic capacity with imperfect CSI, subject to the constraint of the average uplink pilot power. \emph{Thirdly}, the method of Lagrange multipliers is invoked to solve the ergodic rate maximizing problem and thus to obtain the optimal pilot power allocation strategy. The resultant pilot power allocation solution suggests allocating more amount of power to the pilots for estimating the weak reflection channels, which is counter-intuition but effectively improve the ergodic capacity with imperfect CSI. Besides, we also elaborate on the expense of the proposed pilot power allocation strategy upon analyzing the peak-to-average-power-ratio (PAPR) increase quantitatively. \emph{Finally}, the extensive simulation results verify our analysis and reveal some interesting results. For example, for the user in the vicinity of a large IRS, it is suggested to switch off other IRSs and only switch on the IRS nearest the user. On the contrary, for the user near a small IRS, it is better to switch on all IRSs and perform the optimal pilot power allocation for enhancing the achievable rate performance.
\end{abstract}


\IEEEpeerreviewmaketitle

\section{Introduction} 
Recently, intelligent reflecting surface (IRS) \cite{wu2019towards} and its variants, e.g., reconfigurable intelligent surface (RIS) \cite{basar2019wireless, tang2020wireless}, large intelligent surface/antenna (LISA) \cite{liang2019large, hu2018beyond}, have arisen as a promising technology to build the future spectral- and energy-efficient wireless networks in a cost-efficient manner \cite{zhang2020prospective, you2020towards}. Specifically, IRS is a programmable metasurface consisted of a huge number of passive reflecting elements, each of which can alter the incident signals by imposing an independent magnitude and/or phase shift \cite{wu2019towards, di2019smart}. As a result, the propagation environment between transmitters and receivers can be customized according to the specific quality-of-service (QoS) requirements, which brings new degree-of-freedom for enhancing the wireless channel capacity and mitigating the inter-channel interference \cite{wu2020intelligent}. By contrast to the conventional cooperative communications with active relays, the IRS elements passively reflect the incident signals without requiring any radio-frequency (RF) chains \cite{wu2019intelligent, bjornson2019intelligent}. The recollection of the diffused energy in the air makes IRS an excellent option for green communications \cite{han2011green, xu2019sixty}. Besides, IRS is capable of operating in the full-duplex (FD) mode without incurring the self-interference as in the traditional FD relay \cite{wu2019towards, zhang2015full}. Moreover, the low-cost material and patch structure render the IRS can be easily attached to the environment objects, e.g., the walls, ceilings, building surfaces \cite{wu2020intelligent, cui2014coding}. Thanks to these potential benefits, it is expected to deploy massive IRSs in wireless networks to facilitate the future 6G ecosystem \cite{saad2019vision, you2020towards, dang2020should}.

\begin{table*}[!t]
\tiny
\renewcommand\arraystretch{2}
\centering
	\caption{\label{tab1} Comparison of the proposed scheme to existing research results on IRS-assisted systems.}
\begin{tabular}{|l||c||c|c|c|c|c|c|c|c|}
\hline
\textbf{}                   & \textbf{$\star$} & \textbf{\cite{boulogeorgos2020ergodic}-2020} & \textbf{\cite{li2020weighted}-2020} & \textbf{\cite{van2020coverage}-2020} & \textbf{\cite{li2020ergodic}-2020} & \textbf{\cite{guo2019maximum}-2019} & \textbf{\cite{zhou2019robust}-2020} & \textbf{\cite{zhang2019analysis}-2019} & \textbf{\cite{an2020optimal}-2020}\\ \hline
\textbf{Multiple IRSs}      &     $\checkmark$     &     ---    &      $\checkmark$        &       ---       &       ---           &         ---      &         ---       &         $\checkmark$         &   ---                          \\ \hline
\textbf{Ergodic capacity}      &     $\checkmark$      &          $\checkmark$          &        ---          &     $\checkmark$              &         $\checkmark$         &          $\checkmark$         &    ---              &          ---        &         $\checkmark$                                       \\ \hline
\textbf{With imperfect CSI}   &     $\checkmark$      &     ---            &      ---            &           $\checkmark$          &         $\checkmark$         &          $\checkmark$        &         $\checkmark$             &        ---          &       $\checkmark$                                     \\ \hline
\textbf{Pilot power allocation}   &      $\checkmark$     &        ---          &       ---           &          ---        &            ---      &         ---         &       ---           &     ---             &   $\checkmark$                                        \\ \hline
\textbf{IRS selection} &    $\checkmark$       &    ---              &       ---           &   ---               &          ---        &       ---           &     ---             &     ---             &               ---                            \\ \hline
\textbf{System setup}         &     SISO      &        SISO          &       MU-MISO           &       SISO           &       SISO           &      MIMO            &        MU-MISO          &        SISO          &                SISO                                            \\ \hline
\textbf{Phase-shift model}  &     Continuous      &       Continuous           &       Continuous           &         Continuous         &         Discrete         &        Continuous          &     Continuous             &          Continuous        &      Continuous                                              \\ \hline
\end{tabular}\\
\dotfill
\end{table*}

Nevertheless, the integration of IRS into existing wireless networks faces new challenges, one of which is the optimization of the IRS reflection coefficients for facilitating the coherent superposition of dispersed intended signals and/or interference cancellation. Specifically, the optimization of the IRS reflection coefficients is generally coupled with the transmit beamforming design, which is different from the conventional systems without IRS \cite{tse2005fundamentals}. In \cite{wu2019intelligent}, \emph{Wu et al.} considered the IRS-assisted multi-input single-output (MISO) and multi-user MISO systems by formulating the transmit power minimizing problem, subject to the individual signal-to-interference-plus-noise-ratio (SINR) requirements at each user. The formulated problem is then solved by semidefinite relaxation (SDR) and alternating optimization (AO) algorithms to obtain the high-quality approximate solution of the transmit beamforming vectors for multiple users and reflection coefficients at the IRS. Besides, \cite{wu2019intelligent} demonstrated that the IRS behaves quadratic power law with the number of reflecting elements, which is promising for energy-efficient deployment \cite{han2011green}. Soon afterwards, \cite{wu2019beamforming} extended the research results of \cite{wu2019intelligent} upon considering the practical IRS elements with discrete phase shifters. Moreover, \cite{wu2019beamforming} showed that, for the IRS equipped with a large number of reflecting elements, the power loss at the receiver relative to that equipped continuous phase shifters is about $-3.9$ dB for $1$-bit quantized phase shifter and only $-0.9$ dB for $2$-bit quantized phase shifter. Furthermore, \cite{zhang2020capacity} investigated the capacity of IRS-assisted point-to-point multi-input multi-output (MIMO) systems, where the IRS reflection coefficients and transmit precoding are optimized alternatively by assuming that the other is given. Following this, the IRS assisted multi-user and multi-cell MIMO systems are reported in \cite{guo2019weighted} and \cite{pan2020multicell}, respectively, where the rate maximizing problem is solved by AO algorithm and fractional programming technique. The energy-efficiency maximization for IRS assisted multi-user MISO system is investigated in \cite{huang2019reconfigurable}, where the AO algorithm assisted by the gradient descent and majorization-minimization (MM) algorithms are invoked to solve the formulated problem. Recently, the optimization of reflection coefficients at IRS has been performed under the joint consideration with other existing techniques, e.g., millimeter-wave (mmWave) communications \cite{wang2020intelligent}, terahertz (THz) communications \cite{ning2020channel, ma2020intelligent}, visible light communication (VLC) \cite{najafi2019intelligent}, orthogonal frequency division multiplexing (OFDM) \cite{An_arXiv_2021_Reconfigurable}, deep learning (DL) \cite{feng2020deep, taha2019enabling}, unmanned aerial vehicle (UAV) \cite{li2020reconfigurable, ma2020enhancing}, index modulation (IM) \cite{basar2020reconfigurable}, non-orthogonal multiple access (NOMA) \cite{ding2020simple, fu2019intelligent}, physical layer security (PLS) \cite{shen2019secrecy, cui2019secure}, cell-free network \cite{zhang2020joint}, mobile edge computing (MEC) \cite{bai2020latency}, and simultaneous wireless information and power transfer (SWIPT) \cite{wu2019weighted, pan2020intelligent}.

Although significant performance gains offered by IRS have been reported in the open literature, the prior of perfect CSI for all channels involved are generally assumed \cite{wu2019intelligent, wu2019beamforming, zhang2020capacity, guo2019weighted, pan2020multicell, huang2019reconfigurable, An_WCL_2021_The}. In practice, the acquisition of accurate CSI, especially the reflection channel coefficients, is a difficult task due to the passive component at IRS as well as its large number of reflecting elements \cite{wu2020intelligent}. One of the considerations is the semi-passive channel estimation by installing only a small number of sensing elements to estimate the channels of the BS-IRS link and the user-IRS link, respectively \cite{wu2020intelligent, alexandropoulos2020hardware, taha2019enabling}. During the channel estimation phase, the IRS works at the sensing mode by switching off all reflecting elements. The activated sensors receive the pilots from the BS/user and then estimate the channels associated with the sensors. Furthermore, the actual channels of the BS-IRS link and the user-IRS link are obtained by sophisticated signal processing techniques, e.g., compressed sensing, and machine learning \cite{taha2019enabling}. It is noted that the efficiency of the IRS adopting the semi-passive channel estimation approach is heavily dependent on the number of sensors, the quantization level of the analog-to-digital converter (ADC) at the sensors as well as the sensing duration \cite{wu2020intelligent}.

For the extreme implementation without any sensing elements mounted on the IRS, the IRS becomes fully passive \cite{di2019smart, cui2014coding}. Due to the deficiency of baseband signal processing capability, the IRS can no longer estimate the BS-IRS link and the user-IRS link separately. Instead, some recent research endeavour to estimate the cascaded reflection channels, which is the Khatri-Rao product of the BS-IRS link and the user-IRS link \cite{wu2020intelligent, zheng2019intelligent}. Along this line of thought, the most straightforward method is to estimate the direct and cascaded reflection channels in sequence upon switching on only a single element at each time slot, and thus the direct channel and reflection channels can be estimated separately without mutual interference \cite{mishra2019channel}. However, it is obvious to note that the ON/OFF method suffers from severe power loss due to the frequent switching off the IRS. To solve this problem, \cite{jensen2020optimal} proposed a novel channel estimation method by designing orthogonal IRS reflection coefficient vector during the channel estimation phase, thus the cascaded reflection channels can be obtained without the power loss as well as the mutual interference. In addition, \cite{wang2020channel} noted that the reflection channels associated with different users share the common BS-IRS link, and then proposed a novel three-phase channel estimation method, which estimates a set of low-dimensional scaling factors instead of directly estimating reflection channel coefficients. Furthermore, the authors of \cite{you2020channel} reduced the pilot overhead by arranging the IRS elements into groups and estimating only a single channel coefficient for each group, which suffers from the performance penalty. Recently, several new research results have been reported by reducing the dependency of the IRS reflection optimization on accurate CSI. In \cite{han2019large}, the statistic CSI is exploited to perform the phase shift optimization at the expense of system performance. Our prior work has demonstrated that multiple independent random phase shifts can gradually approach the optimal performance, where only the effective sum channel is estimated instead of separately estimating the direct channel and reflection channels \cite{an2020practical}. The interested reader may refer to \cite{wu2020intelligent, gong2020toward, liu2020reconfigurable} and references therein to find the most state-of-the-art channel estimation methods for IRS-assisted wireless systems.

We note that the aforementioned channel estimation methods mainly focus on the pilot design, reflection coefficients optimization and specific estimation algorithms, which are crucial and challenging tasks in IRS-assisted wireless systems \cite{wu2020intelligent, wang2020channel, you2020channel, An_WCL_2021_The}. Nevertheless, there are still some points that can be exploited to further improve the efficiency of the IRS. For example, existing channel estimation methods generally adopt the constant-envelope pilot signals for each user/antenna without considering the pilot power allocation during the channel estimation phase. In \cite{an2020optimal}, we have discussed the pilot power allocation in the single-IRS assisted communication systems and performed the power allocation between the pilots used for estimating the direct link and reflection links. Nonetheless, there exist some limits to the rate performance improvement \cite{an2020optimal}. On one hand, the user gains the performance benefits only in the vicinity of the IRS due to the high path loss, which is the area where the path loss of the reflection channels is significantly severe than that of the direct channel. On the other hand, the IRS is generally equipped with a large number of reflecting elements to compensate for the high path loss. Due to the asymmetry between the direct link and reflection links, the displayed performance improvement in \cite{an2020optimal} is limited, e.g., only effective for a small/moderate number of reflecting elements at the IRS. Furthermore, we note that much of the prior works focus on the communication systems assisted by a single IRS, few of them focus on the design of multiple-IRS assisted communication systems \cite{li2020weighted, zhang2019analysis}.

Against the above background, in this paper, we propose the optimal pilot power allocation strategy for multi-IRS assisted wireless systems. For the sake of exposition, the features of the proposed pilot power allocation scheme are compared to existing research results on IRS-assisted systems in Table \ref{tab1}, where $\star$ denotes the proposed pilot power allocation scheme. More specifically, the contributions of this paper are summarized as follows:
\begin{itemize}
    \item First of all, we introduce a multi-IRS assisted single-input single-output (SISO) communication system considering the practical cases with channel estimation errors, where the most straightforward ON/OFF method is adopted for the sake of brevity. Furthermore, we derive the ergodic achievable rate with imperfect CSI in an explicit closed-form expression, which is dependent on the specific pilot power used for estimating different reflection channels.
    \item Secondly, we formulate the capacity maximizing problem subject to the constraint of the average pilot power to obtain the specific pilot power solution at each time slot. We then solve the problem by the method of Lagrange multipliers based on some effective approximation. The resultant pilot power allocation strategy suggests allocating the same amount of power to the pilots used for estimating the reflection channels via the same IRS while allocating more amount of power to the pilots used for estimating the weak reflection channels. Furthermore, we elaborate on the PAPR increase of pilot signals in the proposed pilot power allocation strategy.
    \item Finally, extensive simulation results verify the derived ergodic capacity and the optimal pilot power allocation strategy. Besides, some interesting results are also summarized. For example, for the user in the vicinity of a large IRS, it is suggested to switch off other IRSs and only switch on the IRS nearest the user; while for the user near a small IRS, it should switch on all IRSs and perform the optimal pilot power allocation. Besides, our simulation also implicitly verify that the IRS should be deployed near the user and distributed IRS deployment can bring extra performance improvements.
\end{itemize}

The rest of this paper is structured as follows. Section \ref{sec2} introduces the system model, while Section \ref{sec3} derives the ergodic capacity with imperfect CSI. Furthermore, Section \ref{sec4} provides the problem formulation, and Section \ref{sec5} solves the formulated problem and gives a general discussion. In Section \ref{sec6}, extensive simulations are provided to verify our analysis. Finally, Section \ref{sec7} concludes the paper.

\emph{Notations:} Scalars are denoted by italic letters, vectors/matrices are denoted by bold-face lower/upper-case letters. ${\left(  \cdot  \right)^T}$, ${\left(  \cdot  \right)^H}$, and ${\left(  \cdot  \right)^ * }$ represent the transpose, Hermitian transpose and conjugate, respectively. ${\mathbb{C}^{x \times y}}$ denotes the space of $x \times y$ complex-valued matrices. For a complex-valued vector ${\bf{x}}$, $\text{diag}\left( {\bf{x}} \right)$ denotes a diagonal matrix with each diagonal element being the corresponding element in ${\bf{x}}$. $\log \left(  \cdot  \right)$ denotes the logarithmic function, while $\mathbb{E}\left\{  \cdot  \right\}$ stands for the expected value. The distribution of a circularly symmetric complex Gaussian (CSCG) random variable with mean $\mu$ and variance ${\sigma ^2}$ is denoted by $\mathcal{CN}\left( {\mu ,{\sigma ^2}} \right)$. $\mathcal{R}\left\{ z \right\}$, $\mathcal{I}\left\{ z \right\}$, and $\left| z \right|$ denote the real part, the imaginary part, and the magnitude of a complex number $z$, respectively.
\begin{figure}[!t]
	\centering
	\includegraphics[width=8.8cm]{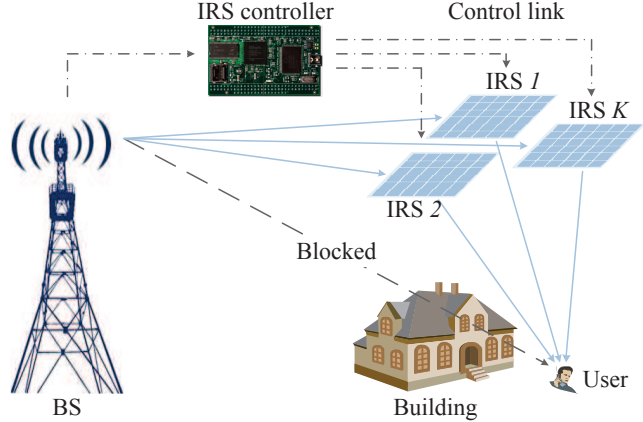}
	\caption{Schematic of a multi-IRS assisted communication system.}
	\label{fig1}
\end{figure}
\section{System Model}\label{sec2}
Let us consider an IRS-assisted downlink communication system, as shown in Fig. \ref{fig1}, where multiple IRSs are deployed at the cell edge to enhance the communication link from the BS to the user. For the sake of brevity, we consider both the BS and user are equipped with a single antenna, and the direct link between the BS and the user is blocked by a local building near the user. We denote the number of IRSs as $K$. The set of the IRSs is denoted by $\mathcal{K} = \left\{ {1, \cdots ,K} \right\}$. Furthermore, we denote the number of reflecting elements at the $k$th IRS as ${M_k},\ \forall k \in \mathcal{K}$, while the corresponding set of reflecting elements is denoted by ${\mathcal{M}_k} = \left\{ {1, \cdots ,{M_k}} \right\},\ \forall k \in \mathcal{K}$. All the IRSs are connected with an IRS controller by the wired/wireless backhaul for feeding back the reflection coefficients. Due to the high path loss, the reflection signals via multiple IRSs is assumed to be below the receiver sensitivity and thus can be ignored.

Furthermore, we assume the quasi-static block-fading model is adopted for all channels involved in Fig. \ref{fig1}. Specifically, the baseband equivalent channels from the BS to the $k$th IRS, and from the $k$th IRS to the user are denoted by ${{\bf{u}}_k} = {\left[ {{u_{k,1}}, \cdots ,{u_{k,{M_k}}}} \right]^T}\in {\mathbb{C}^{{M_k} \times 1}}$, and ${\bf{v}}_k^H = {\left[ {{v_{k,1}}, \cdots ,{v_{k,{M_k}}}} \right]^ * }\in {\mathbb{C}^{1 \times {M_k}}}$, respectively, where ${u_{k,{m_k}}}$ and ${v_{k,{m_k}}}$ denote the channels from the BS and the user, respectively, to the $m_k$th element at the $k$th IRS. Let ${{\bf{\Phi }}_k} = \text{diag}\left( {{\varphi _{k,1}},{\varphi _{k,2}}, \cdots ,{\varphi _{k,{M_k}}}} \right) \in {\mathbb{C}^{{M_k} \times {M_k}}}$ as the reflection coefficients matrix at the $k$th IRS, where ${{\varphi _{k,{m_k}}}}$ denotes the reflection coefficient of the ${m_k}$th element at the $k$th IRS. In this paper, we focus on the phase shift design by assuming the magnitude at each IRS element to be $\left| {{\varphi _{k,{m_k}}}} \right| = 1$ for the maximum reflection power collection\footnote{In fact, the signal reflected by the IRS may experience a slight power loss for the IRS configuration. Besides, the magnitude of reflection coefficients may independently adjusted and/or coupled with the phase shift \cite{wu2020intelligent}. The power allocation considering more practical IRS element model will be left for our future research topic.}. In addition, we assume that the phase shift can be continuously adjusted in $\left[ {0,2\pi } \right)$ to gain essential insight. As a result, the effective channel from the BS to the user is modeled as the superposition of $K$-cluster reflection channels, i.e., $h = \sum\limits_{k = 1}^K {{\bf{v}}_k^H{{\bf{\Phi }}_k}{{\bf{u}}_k}}$.

In this paper, we focus on the narrowband communications. Denote the signal transmitted from the BS over the given band as $s$, which has the normalized power, i.e., ${\left| s \right|^2} = 1$. Therefore, the baseband signal received at the user can be expressed as
\begin{align}\label{eq1}
r = &\left( {\sum\limits_{k = 1}^K {{\bf{v}}_k^H{{\bf{\Phi }}_k}{{\bf{u}}_k}} } \right)\sqrt q s + n \notag\\
 = &\left( {\sum\limits_{k = 1}^K {\sum\limits_{{m_k} = 1}^{{M_k}} {{u_{k,{m_k}}}{\varphi _{k,{m_k}}}v_{k,{m_k}}^*} } } \right)\sqrt q s + n,
\end{align}
where $q$ is the downlink average transmit power at the BS, $n \sim \mathcal{CN}\left( {0,\sigma _n^2} \right)$ is the additive white Gaussian noise (AWGN) with $\sigma _n^2$ denoting the average noise power at the user.

For the link via the ${m_k}$th element at the $k$th IRS, the reflection channel consists of three parts, namely, the BS-IRS link ${u_{k,{m_k}}}$, the IRS phase shift ${\varphi _{k,{m_k}}}$ and the IRS-user link $v_{k,{m_k}}^ * $, where ${\varphi _{k,{m_k}}}$ can be configured by the IRS controller to facilitate the coherent superposition of different reflection links. Nonetheless, the configuration of ${\varphi _{k,{m_k}}}$ needs the prior knowledge of the instantaneous CSI, i.e., ${u_{k,{m_k}}}$, $v_{k,{m_k}}^ * $, which, as stated earlier, is challenging to be obtained separately due to the deficiency of baseband signal processing capability at the IRS. To solve this problem, some novel channel estimation methods have been proposed to estimate the cascaded reflection links instead \cite{mishra2019channel, jensen2020optimal, wang2020channel, you2020channel}. Therefore, (\ref{eq1}) can be rewritten as
\begin{align}
r = \left( {\sum\limits_{k = 1}^K {\sum\limits_{{m_k} = 1}^{{M_k}} {{\varphi _{k,{m_k}}}{h_{k,{m_k}}}} } } \right)\sqrt q s + n,
\end{align}
where we have ${h_{k,{m_k}}} = {u_{k,{m_k}}}v_{k,{m_k}}^ * $ denoting the cascaded reflection channels via the ${m_k}$th element at the $k$th IRS.

Next, we consider the practical channel estimation in the uplink phase, which is crucial for fascinating the coherent superposition of the reflection signals via different IRS elements. We assume the time division duplex (TDD) mode for uplink and downlink transmissions and thus the CSI acquisition in the downlink can be obtained based on the channel reciprocity. Moreover, we adopt the ON/OFF channel method, where the cascaded reflection channels are estimated by only switching on a single reflecting element at each time slot and thus the pilot overhead required is $\sum\nolimits_{k = 1}^K {{M_k}}$. Denote the uplink pilot symbol at the $\left( {\sum\nolimits_{i = 1}^{k - 1} {{M_i}}  + {m_k}} \right)$th time slot as ${x_{k,{m_k}}}$, which is used for estimating the reflection channel via the $m_k$th element at the $k$th IRS upon switching on only the corresponding reflecting element. Denote the normalized pilot symbols at the $\left( {\sum\nolimits_{i = 1}^{k - 1} {{M_i}}  + {m_k}} \right)$th time slot as ${x_{k,{m_k}}}$, i.e., ${\left| {{x_{k,{m_k}}}} \right|^2} = 1$. Therefore, the uplink channel estimation model can be formulated by
\begin{align}\label{eq3}
    {y_{k,{m_k}}} = {\left( {{\varphi _{k,{m_k}}}{h_{k,{m_k}}}} \right)^ * }\sqrt {{p_{k,{m_k}}}} {x_{k,{m_k}}} + {z_{k,{m_k}}},
\end{align}
where ${p_{k,{m_k}}}$ is the uplink pilot power at the $\left( {\sum\nolimits_{i = 1}^{k - 1} {{M_i}}  + {m_k}} \right)$th time slot, ${z_{k,{m_k}}} \sim \mathcal{CN}\left( {0,\sigma _z^2} \right)$ is the AWGN at the BS, ${\varphi _{k,{m_k}}}$ is the IRS reflection coefficient at the $\left( {\sum\nolimits_{i = 1}^{k - 1} {{M_i}}  + {m_k}} \right)$th time slot, which can be set to be an arbitrary complex value satisfying $\left| {{\varphi _{k,{m_k}}}} \right| = 1$.

\emph{Remark 1:} As stated earlier, we note that the ON/OFF method suffers from the power loss due to the fact of frequently switching off the IRS elements. Although recent novel channel estimation methods endeavour to improve the channel estimation accuracy upon optimizing the reflection coefficients at the IRS, they can not be readily applied for the practical implementation with discrete phase shifters at each IRS element. For example, considering the 1-bit quantized phase shifters at each IRS element, it is infeasible to design orthogonal reflection coefficients vectors for the arbitrary number of reflecting elements. Instead, such so-called improved methods suffer from severe mutual interference. In view of this, in this paper, we adopt the ON/OFF method to help us gain insight into the pilot power allocation.

According to (\ref{eq3}), the least squares (LS) estimate of the cascaded reflection channel via the $m_k$th element at the $k$th IRS can be obtained by
\begin{align}
{{\hat h}_{k,{m_k}}} = \frac{1}{{\sqrt {{p_{k,{m_k}}}} }}y_{k,{m_k}}^ * {x_{k,{m_k}}}\varphi _{k,{m_k}}^ * = {h_{k,{m_k}}} + {\varepsilon _{k,{m_k}}},
\end{align}
where ${\varepsilon _{k,{m_k}}} = \frac{1}{{\sqrt {{p_{k,{m_k}}}} }}z_{k,{m_k}}^*{x_{k,{m_k}}}\varphi _{k,{m_k}}^*$ denotes the estimation error with respect to ${h_{k,{m_k}}}$, which is of the distribution ${\varepsilon _{k,{m_k}}} \sim \mathcal{CN}\left( {0,\delta _{k,{m_k}}^2} \right)$ with $\delta _{k,{m_k}}^2 = {{{\sigma _z^2} \mathord{\left/
 {\vphantom {{\sigma _z^2} {{p_{k,{m_k}}}}}} \right.
 \kern-\nulldelimiterspace} {{p_{k,{m_k}}}}}}$ denoting the mean square error (MSE) of the LS channel estimates.

Based on the estimated ${\hat h_{k,{m_k}}}$, the IRS controller will adjust the reflection coefficient of the $m_k$th element at the $k$th IRS. Specifically, the reflection coefficient of the $m_k$th element at the $k$th IRS can be obtained by
\begin{align}\label{eq5}
{\hat \varphi _{k,{m_k}}} = \frac{{\hat h_{k,{m_k}}^*}}{{\left| {{{\hat h}_{k,{m_k}}}} \right|}} = \frac{{h_{k,{m_k}}^* + \varepsilon _{k,{m_k}}^*}}{{\left| {{h_{k,{m_k}}} + {\varepsilon _{k,{m_k}}}} \right|}}.
\end{align}

After optimizing the reflection coefficients and feeding them back to the corresponding IRS, the actual channel between the BS and the user is thus formulated by $h = \sum\nolimits_{k = 1}^K {\sum\nolimits_{{m_k} = 1}^{{M_k}} {{{\hat \varphi }_{k,{m_k}}}{h_{k,{m_k}}}} } $. As a result, the achievable rate of the IRS-assisted SISO system under imperfect CSI can be obtained by 
\begin{align}
R = {\log _2}\left( {1 + \frac{q}{{\sigma _n^2}}{{\left| {\sum\limits_{k = 1}^K {\sum\limits_{{m_k} = 1}^{{M_k}} {{{\hat \varphi }_{k,{m_k}}}{h_{k,{m_k}}}} } } \right|}^2}} \right).
\end{align}
\section{Ergodic Capacity with Imperfect CSI}\label{sec3}
In this section, we will derive the explicit closed-form expression of the ergodic capacity with imperfect CSI. First of all, the ergodic achievable rate can be expressed by \cite{tse2005fundamentals}
\begin{align}\label{eq7}
\mathbb{E}\left\{ R \right\} = \mathbb{E}\left\{ {{{\log }_2}\left( {1 + \text{SNR}} \right)} \right\} \le {\log _2}\left( {1 + \mathbb{E}\left\{ {\text{SNR}} \right\}} \right),
\end{align}
where the average signal-to-noise-ratio (SNR) is defined by
\begin{align}\label{eq8}
    \mathbb{E}\left\{ {\text{SNR}} \right\} = \frac{q}{{\sigma _n^2}}\mathbb{E}\left\{ {{{\left| {\sum\limits_{k = 1}^K {\sum\limits_{{m_k} = 1}^{{M_k}} {{{ \hat \varphi }_{k,{m_k}}}{h_{k,{m_k}}}} } } \right|}^2}} \right\}.
\end{align}

We note that the right hand side (RHS) of (\ref{eq7}) provides a upper bound of the ergodic capacity. In fact, under the high SNR and low SNR, the ergodic achievable rate has the approximate expression of $\mathbb{E}\left\{ R \right\} = {\log _2}\left( {\mathbb{E}\left\{ {\text{SNR}} \right\}} \right)$ and $\mathbb{E}\left\{ R \right\} = {{\mathbb{E}\left\{ {\text{SNR}} \right\}} \mathord{\left/
 {\vphantom {{\mathbb{E}\left\{ {\text{SNR}} \right\}} {\ln 2}}} \right.
 \kern-\nulldelimiterspace} {\ln 2}}$, respectively, both of which are heavily dependent on the average SNR \cite{goldsmith2005wireless}.

Hence, for the sake of brevity, in the following we will focus on the average channel gain by omitting the transmit SNR ${q \mathord{\left/
 {\vphantom {q {\sigma _n^2}}} \right.
 \kern-\nulldelimiterspace} {\sigma _n^2}}$ from (\ref{eq8}). Specifically, upon substituting (\ref{eq5}) into (\ref{eq8}), the average channel gain can be expressed by
\begin{align}\label{eq9}
&\mathbb{E}\left\{ {{{\left| {\sum\limits_{k = 1}^K {\sum\limits_{{m_k} = 1}^{{M_k}} {{{ \hat \varphi } _{k,{m_k}}}{h_{k,{m_k}}}} } } \right|}^2}} \right\} \notag\\
 =& \mathbb{E}\left\{ {{{\left| {\sum\limits_{k = 1}^K {\sum\limits_{{m_k} = 1}^{{M_k}} {\frac{{\left( {h_{k,{m_k}}^ *  + \varepsilon _{k,{m_k}}^ * } \right){h_{k,{m_k}}}}}{{\left| {{h_{k,{m_k}}} + {\varepsilon _{k,{m_k}}}} \right|}}} } } \right|}^2}} \right\} \notag\\
 \overset{\left( a \right)}{=}& \sum\limits_{k = 1}^K {\sum\limits_{{m_k} = 1}^{{M_k}} {\mathbb{E}\left\{ {{{\left| {{h_{k,{m_k}}}} \right|}^2}} \right\}} }  \notag \\
 +& \sum\limits_{k = 1}^K {\sum\limits_{{m_k} = 1}^{{M_k}} {\mathbb{E}\left\{ {\mathcal{R}\left\{ {\frac{{\left( {h_{k,{m_k}}^ *  + \varepsilon _{k,{m_k}}^ * } \right){h_{k,{m_k}}}}}{{\left| {{h_{k,{m_k}}} + {\varepsilon _{k,{m_k}}}} \right|}}} \right\}} \right\}} } \notag\\
&\quad \quad \times \sum\limits_{{{\tilde m}_k} = 1,{{\tilde m}_k} \ne {m_k}}^{{M_k}} {\mathbb{E}\left\{ {\mathcal{R}\left\{ {\frac{{h_{k,{{\tilde m}_k}}^ *  + \varepsilon _{k,{{\tilde m}_k}}^ * {h_{k,{{\tilde m}_k}}}}}{{\left| {{h_{k,{{\tilde m}_k}}} + {\varepsilon _{k,{{\tilde m}_k}}}} \right|}}} \right\}} \right\}} \notag\\
 +& \sum\limits_{k = 1}^K {\sum\limits_{{m_k} = 1}^{{M_k}} {\mathbb{E}\left\{ {\mathcal{I}\left\{ {\frac{{h_{k,{m_k}}^ *  + \varepsilon _{k,{m_k}}^ * {h_{k,{m_k}}}}}{{\left| {{h_{k,{m_k}}} + {\varepsilon _{k,{m_k}}}} \right|}}} \right\}} \right\}} } \notag\\
&\quad \quad \times \sum\limits_{{{\tilde m}_k} = 1,{{\tilde m}_k} \ne {m_k}}^{{M_k}} {\mathbb{E}\left\{ {\mathcal{I}\left\{ {\frac{{h_{k,{{\tilde m}_k}}^ *  + \varepsilon _{k,{{\tilde m}_k}}^ * {h_{k,{{\tilde m}_k}}}}}{{\left| {{h_{k,{{\tilde m}_k}}} + {\varepsilon _{k,{{\tilde m}_k}}}} \right|}}} \right\}} \right\}} \notag\\
 +& \sum\limits_{k = 1}^K {\sum\limits_{{m_k} = 1}^{{M_k}} {\mathbb{E}\left\{ {\mathcal{R}\left\{ {\frac{{\left( {h_{k,{m_k}}^ *  + \varepsilon _{k,{m_k}}^ * } \right){h_{k,{m_k}}}}}{{\left| {{h_{k,{m_k}}} + {\varepsilon _{k,{m_k}}}} \right|}}} \right\}} \right\}} } \notag\\
 & \times \sum\limits_{\tilde k = 1,\tilde k \ne k}^K {\sum\limits_{{m_{\tilde k}} = 1}^{{M_{\tilde k}}} {\mathbb{E}\left\{ {\mathcal{R}\left\{ {\frac{{\left( {h_{\tilde k,{m_{\tilde k}}}^* + \varepsilon _{\tilde k,{m_{\tilde k}}}^*} \right){h_{\tilde k,{m_{\tilde k}}}}}}{{\left| {{h_{\tilde k,{m_{\tilde k}}}} + {\varepsilon _{\tilde k,{m_{\tilde k}}}}} \right|}}} \right\}} \right\}} } \notag\\
 +& \sum\limits_{k = 1}^K {\sum\limits_{{m_k} = 1}^{{M_k}} {\mathbb{E}\left\{ {\mathcal{I}\left\{ {\frac{{\left( {h_{k,{m_k}}^ *  + \varepsilon _{k,{m_k}}^ * } \right){h_{k,{m_k}}}}}{{\left| {{h_{k,{m_k}}} + {\varepsilon _{k,{m_k}}}} \right|}}} \right\}} \right\}} } \notag\\
 &\times \sum\limits_{\tilde k = 1,\tilde k \ne k}^K {\sum\limits_{{m_{\tilde k}} = 1}^{{M_{\tilde k}}} {\mathbb{E}\left\{ {\mathcal{I}\left\{ {\frac{{\left( {h_{\tilde k,{m_{\tilde k}}}^* + \varepsilon _{\tilde k,{m_{\tilde k}}}^*} \right){h_{\tilde k,{m_{\tilde k}}}}}}{{\left| {{h_{\tilde k,{m_{\tilde k}}}} + {\varepsilon _{\tilde k,{m_{\tilde k}}}}} \right|}}} \right\}} \right\}} },
\end{align}
where $\left( a \right)$ holds as the reflection channel ${h_{k,{m_k}}}$ and estimation error ${\varepsilon _{k,{m_k}}}$ associated with different reflecting elements are independent of each other. Furthermore, the second and third terms at the RHS of (\ref{eq9}) reflect the effects of the channel estimation errors from the same IRS, while the fourth and fifth terms reflect the effects of the channel estimation errors from other IRSs.

Furthermore, in the practical implementation, the BS and IRS are generally installed at the high-altitude sites with few local scatters, which means that the BS-IRS link is more likely to form the line-of-sight (LoS) propagation link with high Rician factor and long channel coherence time. Therefore, it is reasonable to assume that the small-scale fading of the cascaded reflection channels is mainly determined by the IRS-user link, i,e., ${h_{k,{m_k}}} \sim \mathcal{CN}\left( {0,\beta  _k^2} \right)$, where $\beta _k^2$ denotes the path loss of the cascaded reflection channels via the $k$th IRS. Note that all the ${M_k}$ elements in the $k$th IRS experience the same degree of path loss. As a result, for the $m_k$th reflecting element at the $k$th IRS, we have $\mathbb{E}\left\{ {{{\left| {{h_{k,{m_k}}}} \right|}^2}} \right\} = \beta _k^2$. In addition, according to the conclusions of \cite{an2020practical}, it is not difficult to verify that, assuming ${h_{k,{m_k}}} \sim \mathcal{CN}\left( {0,\beta _k^2} \right)$ and ${\varepsilon _{k,{m_k}}} \sim \mathcal{CN}\left( {0,\delta _{k,{m_k}}^2} \right)$ for $\forall k \in \mathcal{K},\ \forall {m_k} \in {\mathcal{M}_k}$, we have
\begin{align}
\mathbb{E}\left\{ {\mathcal{R}\left\{ {\frac{{\left( {h_{k,{m_k}}^* + \varepsilon _{k,{m_k}}^*} \right){h_{k,{m_k}}}}}{{\left| {{h_{k,{m_k}}} + {\varepsilon _{k,{m_k}}}} \right|}}} \right\}} \right\} =& \frac{{\sqrt \pi  \beta _k^2}}{{2\sqrt {\beta _k^2 + \delta _{k,{m_k}}^2} }}, \label{eq10}\\
\mathbb{E}\left\{ {\mathcal{I}\left\{ {\frac{{\left( {h_{k,{m_k}}^* + \varepsilon _{k,{m_k}}^*} \right){h_{k,{m_k}}}}}{{\left| {{h_{k,{m_k}}} + {\varepsilon _{k,{m_k}}}} \right|}}} \right\}} \right\} =& 0. \label{eq11}
\end{align}

Upon substituting (\ref{eq10}) and (\ref{eq11}) into (\ref{eq9}), the average channel gain can be further simplified as
\begin{align}
&\mathbb{E}\left\{ {{{\left| {\sum\limits_{k = 1}^K {\sum\limits_{{m_k} = 1}^{{M_k}} {{{\hat \varphi} _{k,{m_k}}}{h_{k,{m_k}}}} } } \right|}^2}} \right\} \notag\\
 =& \sum\limits_{k = 1}^K {\sum\limits_{{m_k} = 1}^{{M_k}} {\beta _k^2} }  \notag\\
 +& \sum\limits_{k = 1}^K {\sum\limits_{{m_k} = 1}^{{M_k}} {\frac{{\sqrt \pi  \beta _k^2}}{{2\sqrt {\beta _k^2 + \delta _{k,{m_k}}^2} }}\sum\limits_{{{\tilde m}_k} = 1,{{\tilde m}_k} \ne {m_k}}^{{M_k}} {\frac{{\sqrt \pi  \beta _k^2}}{{2\sqrt {\beta _k^2 + \delta _{k,{{\tilde m}_k}}^2} }}} } }  \notag\\
 +& \sum\limits_{k = 1}^K {\sum\limits_{{m_k} = 1}^{{M_k}} {\frac{{\sqrt \pi  \beta _k^2}}{{2\sqrt {\beta _k^2 + \delta _{k,{m_k}}^2} }}\sum\limits_{\tilde k = 1,\tilde k \ne k}^K {\sum\limits_{{m_{\tilde k}} = 1}^{{M_{\tilde k}}} {\frac{{\sqrt \pi  \beta _{\tilde k}^2}}{{2\sqrt {\beta _{\tilde k}^2 + \delta _{\tilde k,{m_{\tilde k}}}^2} }}} } } }  \notag\\
 =& \sum\limits_{k = 1}^K {{M_k}\beta _k^2}  \notag\\
 +&\frac{\pi }{4}\sum\limits_{k = 1}^K {\beta _k^4\sum\limits_{{m_k} = 1}^{{M_k}} {\frac{1}{{\sqrt {\beta _k^2 + \frac{{\sigma _z^2}}{{{p_{k,{m_k}}}}}} }}\sum\limits_{{{\tilde m}_k} = 1,{{\tilde m}_k} \ne {m_k}}^{{M_k}} {\frac{1}{{\sqrt {\beta _k^2 + \frac{{\sigma _z^2}}{{{p_{k,{{\tilde m}_k}}}}}} }}} } } \notag \\
 +& \frac{\pi }{4}\sum\limits_{k = 1}^K {\sum\limits_{{m_k} = 1}^{{M_k}} {\frac{{\beta _k^2}}{{\sqrt {\beta _k^2 + \frac{{\sigma _z^2}}{{{p_{k,{m_k}}}}}} }}\sum\limits_{\tilde k = 1,\tilde k \ne k}^K {\sum\limits_{{m_{\tilde k}} = 1}^{{M_{\tilde k}}} {\frac{{\beta _{\tilde k}^2}}{{\sqrt {\beta _{\tilde k}^2 + \frac{{\sigma _z^2}}{{{p_{\tilde k,{m_{\tilde k}}}}}}} }}} } } }, \label{eq12}
\end{align}
which will be further reduced to
\begin{align}\label{eq12-1}
\mathbb{E}\left\{ {{{\left| h \right|}^2}} \right\} =& \sum\limits_{k = 1}^K {{M_k}\beta _k^2}  + \frac{\pi }{4}\sum\limits_{k = 1}^K {\beta _k^2{M_k}\left( {{M_k} - 1} \right)} \notag \\
 +& \frac{\pi }{4}\sum\limits_{k = 1}^K {{M_k}{\beta _k}\sum\limits_{\tilde k = 1,\tilde k \ne k}^K {{M_{\tilde k}}{\beta _{\tilde k}}} },
\end{align}
if there are no channel estimation errors. It is not difficult to verify that (\ref{eq12-1}) reveals the quadratic power law with the number of reflecting elements, which is consistent with the conclusions of \cite{wu2019intelligent, liu2020reconfigurable}.

Upon substituting (\ref{eq12}) into (\ref{eq7}), the approximate ergodic channel capacity can be obtained, which is omitted here for the sake of brevity. We note that in (\ref{eq12}), the average channel gain is dependent on the power of pilots used for estimating different reflection channels. Therefore, we desire to find the optimal pilot power strategy that maximizes the ergodic achievable rate with imperfect CSI, which will be addressed in the next section.

\section{Problem Formulation}\label{sec4}
In this section, we will provide the problem formulation. More explicitly, we endeavour to optimize the pilot power for maximizing the ergodic channel capacity with imperfect CSI. Based on this consideration, the optimization problem can be formulated by
\begin{align}\label{eq13}
\begin{array}{*{20}{l}}
{\left( {P1} \right)}&{\mathop {\max }\limits_\mathcal{P} }&{\mathbb{E}\left\{ R \right\}}\\
{}&{s.t.}&{\sum\limits_{k = 1}^K {\sum\limits_{{m_k} = 1}^{{M_k}} {{p_{k,{m_k}}}} }  = \sum\limits_{k = 1}^K {{M_k}} p},
\end{array}
\end{align}
where $p$ is the average uplink pilot power at the user; while
\begin{align}
    \mathcal{P} = \left\{ {{p_{1,1}}, \cdots ,{p_{1,{M_1}}}, \cdots ,{p_{K,1}}, \cdots ,{p_{K,{M_K}}}} \right\}
\end{align}
is the set of the specific pilot power for entire $\sum\nolimits_{k = 1}^K {{M_k}} $ time slots during the channel estimation phase.

Upon substituting (\ref{eq12}) into (\ref{eq13}) and removing the constant terms and positive coefficient, the problem $\left( {P1} \right)$ in (\ref{eq13}) can be reduced to
\begin{align}\label{eq14}
\begin{array}{*{20}{l}}
{\left( {P2} \right)}&{\mathop {\max }\limits_\mathcal{P} }&{\varphi \left( \mathcal{P} \right)}\\
{}&{s.t.}&{\sum\limits_{k = 1}^K {\sum\limits_{{m_k} = 1}^{{M_k}} {{p_{k,{m_k}}}} }  = \sum\limits_{k = 1}^K {{M_k}} p},
\end{array}
\end{align}
where the equivalent objective function $\varphi \left( \mathcal{P} \right)$ is defined by
\begin{align}
\varphi \left( \mathcal{P} \right) =& \sum\limits_{k = 1}^K {\beta _k^4\sum\limits_{{m_k} = 1}^{{M_k}} {\frac{1}{{\sqrt {\beta _k^2 + \frac{{\sigma _z^2}}{{{p_{k,{m_k}}}}}} }}} \sum\limits_{{{\tilde m}_k} = 1,{{\tilde m}_k} \ne {m_k}}^{{M_k}} {\frac{1}{{\sqrt {\beta _k^2 + \frac{{\sigma _z^2}}{{{p_{k,{{\tilde m}_k}}}}}} }}} } \notag \\
 +& \sum\limits_{k = 1}^K {\sum\limits_{{m_k} = 1}^{{M_k}} {\frac{{\beta _k^2}}{{\sqrt {\beta _k^2 + \frac{{\sigma _z^2}}{{{p_{k,{m_k}}}}}} }}} } \sum\limits_{\tilde k = 1,\tilde k \ne k}^K {\sum\limits_{{m_{\tilde k}} = 1}^{{M_{\tilde k}}} {\frac{{\beta _{\tilde k}^2}}{{\sqrt {\beta _{\tilde k}^2 + \frac{{\sigma _z^2}}{{{p_{\tilde k,{m_{\tilde k}}}}}}} }}} }.
\end{align}
\section{The Proposed Optimal Pilot Power Allocation Strategy}\label{sec5}
In this section, we will use the method of Lagrange multipliers to obtain the solution of the problem $\left( {P2} \right)$. First of all, the Lagrangian function $f\left( {\mathcal{P},\lambda } \right)$ is defined as follows
\begin{align}
f\left( {\mathcal{P},\lambda } \right) = \varphi \left( \mathcal{P} \right) - \lambda \left( {\sum\limits_{k = 1}^K {\sum\limits_{{m_k} = 1}^{{M_k}} {{p_{k,{m_k}}}} }  - \sum\limits_{k = 1}^K {{M_k}} p} \right),
\end{align}
where $\lambda $ is the Lagrange multiplier.

Next, we proceed to find the stationary point of the Lagrange function upon yielding the first-order partial derivatives of $f\left( {\mathcal{P},\lambda } \right)$ to be zero. More specifically, for the $m_k$th element at the $k$th IRS, we have
\begin{align}\label{eq17}
\frac{{\partial f}}{{\partial {p_{k,{m_k}}}}} =& \frac{{\beta _k^4\sigma _z^2}}{{p_{k,{m_k}}^2\sqrt {{{\left( {\beta _k^2 + \frac{{\sigma _z^2}}{{{p_{k,{m_k}}}}}} \right)}^3}} }}\sum\limits_{{{\tilde m}_k} = 1,{{\tilde m}_k} \ne {m_k}}^{{M_k}} {\frac{1}{{\sqrt {\beta _k^2 + \frac{{\sigma _z^2}}{{{p_{k,{{\tilde m}_k}}}}}} }}} \notag \\
 +& \frac{{\beta _k^2\sigma _z^2}}{{p_{k,{m_k}}^2\sqrt {{{\left( {\beta _k^2 + \frac{{\sigma _z^2}}{{{p_{k,{m_k}}}}}} \right)}^3}} }}\sum\limits_{\tilde k = 1,\tilde k \ne k}^K {\sum\limits_{{m_{\tilde k}} = 1}^{{M_{\tilde k}}} {\frac{{\beta _{\tilde k}^2}}{{\sqrt {\beta _{\tilde k}^2 + \frac{{\sigma _z^2}}{{{p_{\tilde k,{m_{\tilde k}}}}}}} }}} } \notag \\ 
 -& \lambda =0.
\end{align}

Furthermore, let us first focus on the set of the first-order partial derivatives associated with the $k$th IRS, i.e., $\left\{ {{{\partial f} \mathord{\left/
 {\vphantom {{\partial f} {\partial {p_{k,1}}}}} \right.
 \kern-\nulldelimiterspace} {\partial {p_{k,1}}}}, \cdots ,{{\partial f} \mathord{\left/
 {\vphantom {{\partial f} {\partial {p_{k,{M_k}}}}}} \right.
 \kern-\nulldelimiterspace} {\partial {p_{k,{M_k}}}}}} \right\}$. Upon comparing arbitrary pairs of $\left( {{m_k} \in {\mathcal{M}_k},\ {{m'}_k} \in {\mathcal{M}_k}} \right)$, we have
\begin{align}\label{eq18}
&\frac{{\beta _k^2}}{{p_{k,{m_k}}^2\sqrt {{{\left( {\beta _k^2 + \frac{{\sigma _z^2}}{{{p_{k,{m_k}}}}}} \right)}^3}} }}\sum\limits_{{{\tilde m}_k} = 1,{{\tilde m}_k} \ne {m_k}}^{{M_k}} {\frac{1}{{\sqrt {\beta _k^2 + \frac{{\sigma _z^2}}{{{p_{k,{{\tilde m}_k}}}}}} }}} \notag\\
 +& \frac{1}{{p_{k,{m_k}}^2\sqrt {{{\left( {\beta _k^2 + \frac{{\sigma _z^2}}{{{p_{k,{m_k}}}}}} \right)}^3}} }}\sum\limits_{\tilde k = 1,\tilde k \ne k}^K {\sum\limits_{{m_{\tilde k}} = 1}^{{M_{\tilde k}}} {\frac{{\beta _{\tilde k}^2}}{{\sqrt {\beta _{\tilde k}^2 + \frac{{\sigma _z^2}}{{{p_{\tilde k,{m_{\tilde k}}}}}}} }}} }  \notag\\
 =& \frac{{\beta _k^2}}{{p_{k,{{m'}_k}}^2\sqrt {{{\left( {\beta _k^2 + \frac{{\sigma _z^2}}{{{p_{k,{{m'}_k}}}}}} \right)}^3}} }}\sum\limits_{{{\tilde m}_k} = 1,{{\tilde m}_k} \ne {{m'}_k}}^{{M_k}} {\frac{1}{{\sqrt {\beta _k^2 + \frac{{\sigma _z^2}}{{{p_{k,{{\tilde m}_k}}}}}} }}}  \notag\\
 +& \frac{1}{{p_{k,{{m'}_k}}^2\sqrt {{{\left( {\beta _k^2 + \frac{{\sigma _z^2}}{{{p_{k,{{m'}_k}}}}}} \right)}^3}} }}\sum\limits_{\tilde k = 1,\tilde k \ne k}^K {\sum\limits_{{m_{\tilde k}} = 1}^{{M_{\tilde k}}} {\frac{{\beta _{\tilde k}^2}}{{\sqrt {\beta _{\tilde k}^2 + \frac{{\sigma _z^2}}{{{p_{\tilde k,{m_{\tilde k}}}}}}} }}} }.
\end{align}

We note that it is hard to obtain the solution of (\ref{eq18}) since ${M_k}$ variables, i.e., $\left\{ {{p_{k,1}}, \cdots ,{p_{k,{M_k}}}} \right\}$ in (\ref{eq18}) are coupled with each other. Upon constructing the same terms on both sides of (\ref{eq18}), we have
\begin{align}\label{eq19}
&\frac{{\beta _k^2\left( {{A_k} - {1 \mathord{\left/
 {\vphantom {1 {\sqrt {\beta _k^2 + \frac{{\sigma _z^2}}{{{p_{k,{m_k}}}}}} }}} \right.
 \kern-\nulldelimiterspace} {\sqrt {\beta _k^2 + \frac{{\sigma _z^2}}{{{p_{k,{m_k}}}}}} }}} \right)}}{{p_{k,{m_k}}^2\sqrt {{{\left( {\beta _k^2 + \frac{{\sigma _z^2}}{{{p_{k,{m_k}}}}}} \right)}^3}} }} + \frac{{{B_k}}}{{p_{k,{m_k}}^2\sqrt {{{\left( {\beta _k^2 + \frac{{\sigma _z^2}}{{{p_{k,{m_k}}}}}} \right)}^3}} }} \notag\\
 = &\frac{{\beta _k^2\left( {{A_k} - {1 \mathord{\left/
 {\vphantom {1 {\sqrt {\beta _k^2 + \frac{{\sigma _z^2}}{{{p_{k,{{m'}_k}}}}}} }}} \right.
 \kern-\nulldelimiterspace} {\sqrt {\beta _k^2 + \frac{{\sigma _z^2}}{{{p_{k,{{m'}_k}}}}}} }}} \right)}}{{p_{k,{{m'}_k}}^2\sqrt {{{\left( {\beta _k^2 + \frac{{\sigma _z^2}}{{{p_{k,{{m'}_k}}}}}} \right)}^3}} }} + \frac{{{B_k}}}{{p_{k,{{m'}_k}}^2\sqrt {{{\left( {\beta _k^2 + \frac{{\sigma _z^2}}{{{p_{k,{{m'}_k}}}}}} \right)}^3}} }},
\end{align}
where ${A_k}$ and ${B_k}$ are defined by
\begin{align}
{A_k} = \sum\limits_{{m_k} = 1}^{{M_k}} {\frac{1}{{\sqrt {\beta _k^2 + \frac{{\sigma _z^2}}{{{p_{k,{m_k}}}}}} }}},\
{B_k} = \sum\limits_{\tilde k = 1,\tilde k \ne k}^K {\sum\limits_{{m_{\tilde k}} = 1}^{{M_{\tilde k}}} {\frac{{\beta _{\tilde k}^2}}{{\sqrt {\beta _{\tilde k}^2 + \frac{{\sigma _z^2}}{{{p_{\tilde k,{m_{\tilde k}}}}}}} }}} },
\end{align}
which are independent of ${m_k}$ and ${m'_k}$.

According to (\ref{eq19}), we can easily obtain that ${p_{k,{m_k}}} = {p_{k,{{m'}_k}}}$ for $\forall {m_k} \in {\mathcal{M}_k},\ \forall {m'_k} \in {\mathcal{M}_k}$. As a result, the pilot power for estimating the cascaded reflection channels via the same IRS can be obtained by
\begin{align}\label{eq22}
{p_{k,1}} =  \cdots = {p_{k,{M_k}}} = {p_k},\quad \forall k \in \mathcal{K},
\end{align}
where ${p_k}$ is the average pilot power allocated for the $k$th IRS. (\ref{eq22}) means that one should allocate identical amount of power to the pilots for estimating the channel coefficients of the reflection link via the same IRS, which is consistent with our intuition, since they generally experience the same large-scale fading. In the next, we will focus on the pilot power allocation among different IRSs having different path loss.
 
More explicitly, upon substituting (\ref{eq22}) into (\ref{eq17}) and simplifying, we have
\begin{align}\label{eq23}
\frac{{\partial f}}{{\partial {p_k}}} =& \frac{{\beta _k^2\sigma _z^2}}{{p_k^2\sqrt {{{\left( {\beta _k^2 + \frac{{\sigma _z^2}}{{{p_k}}}} \right)}^3}} }}\sum\limits_{\tilde k = 1}^K {\frac{{\beta _{\tilde k}^2{M_{\tilde k}}}}{{\sqrt {\left( {\beta _{\tilde k}^2 + \frac{{\sigma _z^2}}{{{p_{\tilde k}}}}} \right)} }}} \notag \\
 -& \frac{{\beta _k^4\sigma _z^2}}{{p_k^2{{\left( {\beta _k^2 + \frac{{\sigma _z^2}}{{{p_k}}}} \right)}^2}}} - \lambda  = 0,\quad k \in \mathcal{K}.
\end{align}

Nevertheless, it is not trivial to solve the equation set with $K$ equations in (\ref{eq23}), since (\ref{eq23}) is actually an equation of higher degree with respect to $\left\{ {{p_1}, \cdots ,{p_K}} \right\}$, which are coupled with each other. Therefore, in the next, we will find a high-quality approximate solution instead. Considering that for the practical implementation, one may expect that the communication systems operate at a moderate SNR region for meeting the specific QoS requirement, i.e., ${{{p_k}\beta _k^2} \mathord{\left/
 {\vphantom {{{p_k}\beta _k^2} {\sigma _z^2}}} \right.
 \kern-\nulldelimiterspace} {\sigma _z^2}} \ge \gamma $, where $\gamma $ is the minimum SNR requirement. Therefore, it is a reasonable assumption that $\beta _k^2 + {{\sigma _z^2} \mathord{\left/
 {\vphantom {{\sigma _z^2} {{p_k}}}} \right.
 \kern-\nulldelimiterspace} {{p_k}}} \approx \beta _k^2,\ \forall k \in \mathcal{K}$. Based on such an approximation, (\ref{eq23}) is reduced to
 \begin{align}
\frac{{\sigma _z^2}}{{p_k^2{\beta _k}}}\sum\limits_{\tilde k = 1}^K {{\beta _{\tilde k}}{M_{\tilde k}}}  - \frac{{\sigma _z^2}}{{p_k^2}} - \lambda  = 0.
 \end{align}
 
Furthermore, recalling that the average pilot power constraint of $\sum\limits_{k = 1}^K {{p_k}{M_k}}  = \sum\limits_{k = 1}^K {{M_k}} p$, the optimal pilot power allocation strategy can be obtained by
\begin{align}\label{eq27}
{p_k} = \frac{{\sqrt {\left( {\frac{1}{{{\beta _k}}}\sum\limits_{\hat k = 1}^K {{\beta _{\hat k}}{M_{\hat k}}}  - 1} \right)} }}{{\sum\limits_{\tilde k = 1}^K {{M_{\tilde k}}\sqrt {\left( {\frac{1}{{{\beta _{\tilde k}}}}\sum\limits_{\hat k = 1}^K {{\beta _{\hat k}}{M_{\hat k}}}  - 1} \right)} } }}\sum\limits_{\tilde k = 1}^K {{M_{\tilde k}}} p.
\end{align}
\begin{figure}[!t]
	\centering
	\includegraphics[width=8.8cm]{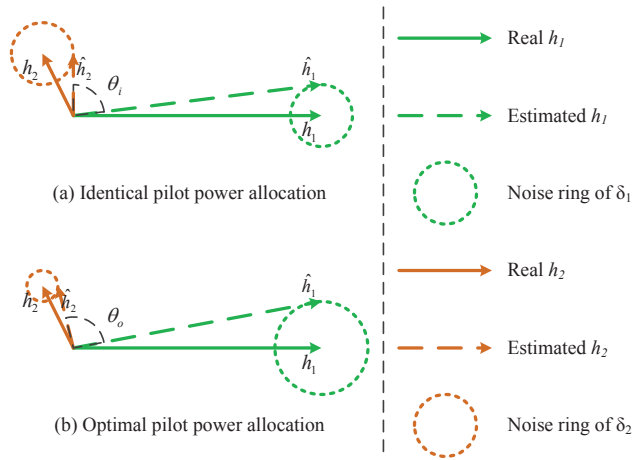}
	\caption{Comparison of two pilot power allocation schemes, where we have $K = 2$, ${M_1} = {M_2}=1$, and $\beta _1^2 = 16\beta _2^2$. It is noted that the proposed optimal pilot power allocation strategy is capable of improving the accuracy of the phase shift in the statistical sense.}
	\label{fig2}
\end{figure}

In the practical deployment, each IRS is generally equipped a large number of reflecting elements to compensate for the high path loss. Therefore, we have $\frac{1}{{{\beta _k}}}\sum\limits_{\hat k = 1}^K {{\beta _{\hat k}}{M_{\hat k}}}  \gg 1$ for $\forall k \in \mathcal{K}$. Accordingly, the power allocation strategy in (\ref{eq27}) are further simplified to
\begin{align}\label{eq28}
{p_k} = \frac{{\sum\limits_{\tilde k = 1}^K {{M_{\tilde k}}} p}}{{\sqrt {{\beta _k}} \sum\limits_{\tilde k = 1}^K {\frac{{{M_{\tilde k}}}}{{\sqrt {{\beta _{\tilde k}}} }}} }},
\end{align}
which is the important conclusion in this paper. The conclusion of (\ref{eq28}) suggests that one should allocate more amount of power to obtain a more accurate estimate of the weak reflection channels, which is a little counter-intuition. Next, we will consider two special cases to help us understand this conclusion more deeply.
\subsection{${M_1}  =  \cdots  = {M_K}$}
Let us first consider the case that all IRSs are equipped with the same number of reflecting elements, i.e., ${M_1} =  \cdots  = {M_K}$. Hence, the optimal pilot power allocation strategy in (\ref{eq28}) is reduced to
\begin{align}\label{eq28-1}
{p_k} = \frac{{Kp}}{{\sqrt {{\beta _k}} \sum\limits_{\tilde k = 1}^K {\frac{1}{{\sqrt {{\beta _{\tilde k}}} }}} }},
\end{align}
which characterizes the inverse square-root pilot power scalling law with respect to the path loss of the corresponding reflection channel, i.e., ${p_k} \propto {1 \mathord{\left/
 {\vphantom {1 {\sqrt {{\beta _k}} }}} \right.
 \kern-\nulldelimiterspace} {\sqrt {{\beta _k}} }}$.
 
More explicitly, we give a simple example to explain the conclusions in (\ref{eq28-1}). For the sake of elaboration, we consider a communication system assisted by two IRSs, each of which is equipped with only a single reflecting element. The specific channel coefficients are shown in Fig. \ref{fig2}, where the statistical channel gain of such two links are set to be ${\beta _1^2} = 16{\beta _2^2}$. It should be noted that the phase diffusion of the channel estimate of the weak channel is more severe than that for the strong channel. Therefore, estimating a more accurate estimate of the weak channel at the expense of a more wore estimate of the strong channel is a compromise option. As can be seen from Fig. \ref{fig2}, the proposed optimal pilot power allocation scheme is capable of improving the accuracy of the IRS phase shift in the statistical sense.

\subsection{${\beta _1} =  \cdots  = {\beta _K}$}
Next we consider the case that all IRSs experience the same degree of path loss, i.e., ${\beta _1}  =  \cdots  = {\beta _K}$. Hence, the optimal pilot power allocation strategy in (\ref{eq28}) is reduced to ${p_1}  =  \cdots  = {p_K} =p$, which means the proposed pilot power allocation strategy is reduced to the identical pilot power allocation strategy. In fact, when multiple IRSs experience the same degree of path loss, it is noted that distributed IRS deployment is equivalent to the localized IRS deployment with $\sum\nolimits_{k = 1}^K {{M_k}}$ reflecting elements. In summary, the proposed pilot power allocation strategy operates effectively in the distributed deployment of IRSs with different path loss.

\emph{Remark 2:} The optimal pilot power allocation scheme improves the ergodic achievable rate with imperfect CSI. Nevertheless, the performance improvement is achieved at the expense of the PAPR increase. More specifically, the PAPR of the pilot symbols in the optimal pilot power allocation strategy is defined by \cite{goldsmith2005wireless}
\begin{align}
{\text{PAPR}_o} = \frac{{\max \left\{ {{p_k}} \right\}}}{{\mathbb{E}\left\{ {{p_k}} \right\}}} = \frac{{{p_{\max }}}}{p},
\end{align}
where ${{p_{\max }}}$ is defined as (\ref{eq28}) upon substituting ${\beta _{\min }}$ instead of ${\beta _k}$. Compared to the constant-envelope pilot symbols in the identical pilot power allocation scheme, the optimal pilot power allocation scheme will cause the PAPR increase. More explicitly, the PAPR increase is quantitatively constrained by
\begin{align}
    \sqrt {\frac{{{\beta _{\max }}}}{{{\beta _{\min }}}}}  \ge {\text{PAPR}_o} \ge {\text{PAPR}_i} = 0\ dB,
\end{align}
which can be easily obtained by the scaling operation on (\ref{eq28}), ${\beta _{\max }}$ and ${\beta _{\min }}$ denote the maximum and minimum value of $\left\{ {{\beta _1}, \cdots ,{\beta _K}} \right\}$, respectively. We note that when ${\beta _{\max }} = {\beta _{\min }}$, we have ${\text{PAPR}_o} = 0$ dB, which is consistent with our prior analysis.

\section{Simulation Results}\label{sec6}
In this section, we will provide simulation results to verify our analysis. The average noise power at the BS and user are set to be $\sigma _z^2 =  - 110$ dBm and $\sigma _n^2 =  - 90$ dBm, respectively. The average downlink transmit power at the BS is set to be $q = 40$ dBm. In our simulations, the location of the BS is set to be $\left( {0,0,{d_h}} \right)$ with the height of ${d_h} = 10$ m. We consider that the direct link between the BS and user is blocked throughout the user's movement by assuming that the user is in the vicinity of the building, as shown in Fig. \ref{fig1}. The path loss of the BS-IRS and IRS-user links is modeled by ${\beta ^2} = {C_0}{d^{ - \alpha }}$, where ${C_0} =  - 20$ dB is the path loss at the reference distance of ${d_r} = 1$ m. The path loss exponents of the BS-IRS and IRS-user links are set to be ${\alpha _{bi}} = 2.2$ and ${\alpha _{iu}} = 2.8$, respectively \cite{zhang2020capacity}. Furthermore, the Rician fading channel model is adopted for both the BS-IRS link and the IRS-user link, with the Rician factors of ${K_{bi}} = \infty $ and ${K_{iu}} = 0$, respectively. All simulation results are obtained by averaging $1000$ channel realizations. For the sake of brevity, other specific parameters will be specified in the following figures.

\begin{figure}[!t]
	\centering
	\includegraphics[width=8.8cm]{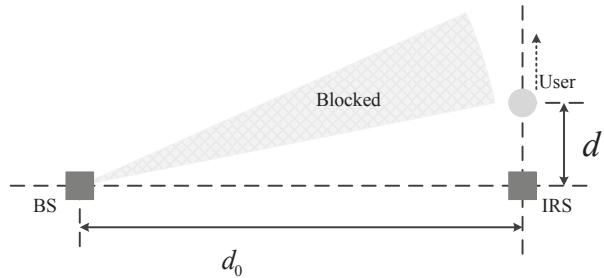}
	\caption{The simulation setup of the single-IRS assisted communication system (from the top view).}
	\label{fig3}
\end{figure}\begin{figure}[!t]
	\centering
	\includegraphics[width=8.8cm]{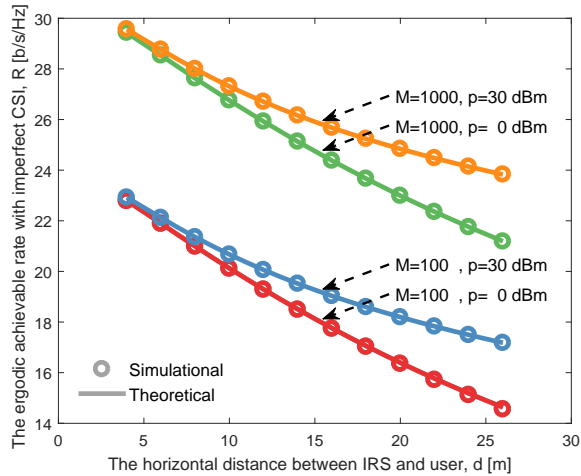}
	\caption{The ergodic achievable rate versus the horizontal distance between the IRS and user.}
	\label{fig4}
\end{figure}
First, let us verify the ergodic capacity performance with imperfect CSI. Specifically, we consider a simple simulation setting as shown in Fig. \ref{fig3}, where an IRS is deployed at the location $\left( {{d_0},0,{d_h}} \right)$, with the horizontal BS-IRS distance of ${d_0} = 50$ m. The multiple-IRS scenarios will be verified later by the proposed pilot power allocation scheme. We consider that the user moves far away from the IRS along the line of $\left( {{d_0},d,0} \right)$, where the horizontal distance between the IRS and user is increased from $d = 4$ m to $d = 26$ m. Fig. \ref{fig4} plots the ergodic achievable rate under different setups, considering the number of reflecting elements of $M = 100$ and $M = 1000$, respectively; the uplink average pilot power of $p = 0$ dBm and $p = 30$ dBm, respectively. It can be seen from Fig. \ref{fig4} that the achievable rate will be improved with the increase of the number of reflecting elements or the average pilot power. In all cases, our theoretical analysis is matched with the simulation results perfectly. In addition, one may note that, as the number of reflecting elements increases from $M = 100$ to $M = 1000$, the ergodic capacity is improved by about $6.6$ b/s/Hz, which is consistent with the quadratic power law in \cite{wu2019intelligent}, i.e., $2{\log _2}10 \approx 6.6$ b/s/Hz. Furthermore, we note that for the same number of reflecting elements, the performance improvement caused by the increased average pilot power becomes significant as the user moves far away from the IRS. For example, the rate improvement is about $3$ b/s/Hz when increasing the average pilot power from $p = 0$ dBm to $p = 30$ dBm for the user at the location of $d = 26$ m. Since the path loss becomes severe as the user moves far away from the IRS, it is suggested to increase the average pilot power to obtain better channel estimates and thus facilitate the coherent superposition between multiple reflection channels for the user far away from the IRS.

\begin{figure}[!t]
	\centering
	\includegraphics[width=8.8cm]{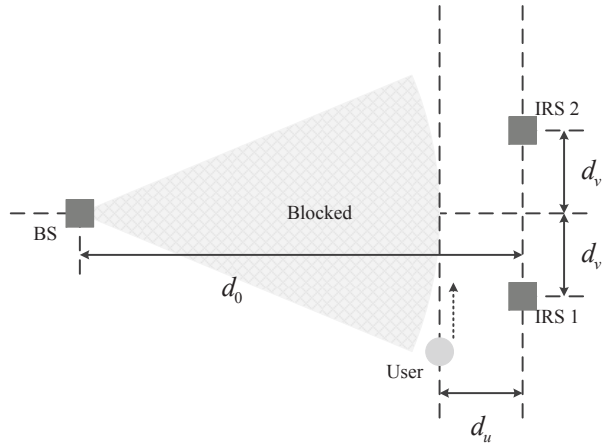}
	\caption{The simulation setup of the two-IRS assisted communication system (from the top view).}
	\label{fig5}
\end{figure}

\begin{figure}[!t]
	\centering
	\includegraphics[width=8.8cm]{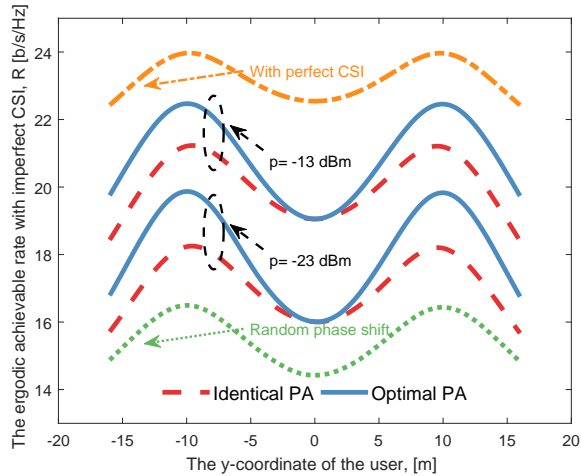}
	\caption{The ergodic achievable rate versus the y-coordinate of the user (symmetrical arrangement of two IRSs).}
	\label{fig6}
\end{figure}

Next, we will consider a communication system, as shown in Fig. \ref{fig5}, with two distributed IRSs to elaborate on the proposed pilot power allocation strategy in detail. IRS-$1$ and IRS-$2$ are located at $\left( {{d_0}, - {d_v},{d_h}} \right)$ and $\left( {{d_0},{d_v},{d_h}} \right)$, respectively, where ${d_v}$ is set to be ${d_v} = 10$ m. The user is located at the line parallel to the line connecting IRS-$1$ and IRS-$2$, with the horizontal distance of ${d_u} = 2$ m. As a result, the user's location is denoted by $\left( {{d_0} - {d_u},d,0} \right)$. The number of reflecting elements at IRS-$1$ and IRS-$2$ are set to be ${M_1} = {M_2} = 100$. We consider the y-coordinate of the user is increased from $d =  - 16$ m to $d =  16$ m. The ergodic capacity comparison of different power allocation strategies is shown in Fig. \ref{fig6}. For the sake of elaboration, the optimal ergodic capacity with perfect CSI and the worst ergodic capacity with random phase shift configuration are also plotted in Fig. \ref{fig6}. Specifically, we consider two cases of the average pilot power of $p =  - 13$ dBm and $p =  - 23$ dBm, respectively. On one hand, the increased average pilot power brings performance improvement, which is consistent with our intuition. On the other hand, the proposed pilot power allocation strategy further widens the performance gap with the identical pilot power allocation strategy, especially for the user in the vicinity of an arbitrary one of two IRSs. This uplifting result is consistent with our theoretical analysis, which means that the ergodic capacity with imperfect CSI can be improved upon finely allocating a different amount of power to the pilots used for estimating different reflection channels.

\begin{figure}[!t]
	\centering
	\includegraphics[width=8.8cm]{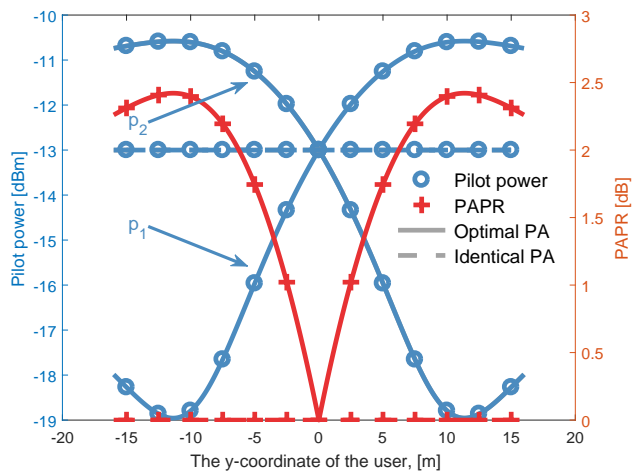}
	\caption{The power and PAPR of pilot signals versus the y-coordinate of the user (symmetrical arrangement of two IRSs).}
	\label{fig7}
\end{figure}
Moreover, the specific pilot power used for estimating the reflection channels of the BS--IRS-$1$--user link and the BS--IRS-$2$--user link are shown as the left y-axis in Fig. \ref{fig7}, where we consider the average pilot power of $p =  - 13$ dBm. It can be seen from Fig. \ref{fig7} that, when the user moves in the vicinity of the IRS-$1$, the proposed pilot power allocation strategy suggests allocating more amount of power to the pilots for estimating the reflection channels via the IRS-$2$, which is counter-intuitive but effectively improves the achievable rate of the user near the IRS-$1$. The same conclusions can be summarized as the user moves near the IRS-$2$. In addition, the PAPR comparison of different pilot power allocation strategies is shown in Fig. \ref{fig7}, which corresponds to the right y-axis. It can be seen from Fig. \ref{fig7} that the proposed pilot power allocation strategy suffers from the PAPR increase.\begin{figure}[!t]
	\centering
	\includegraphics[width=8.8cm]{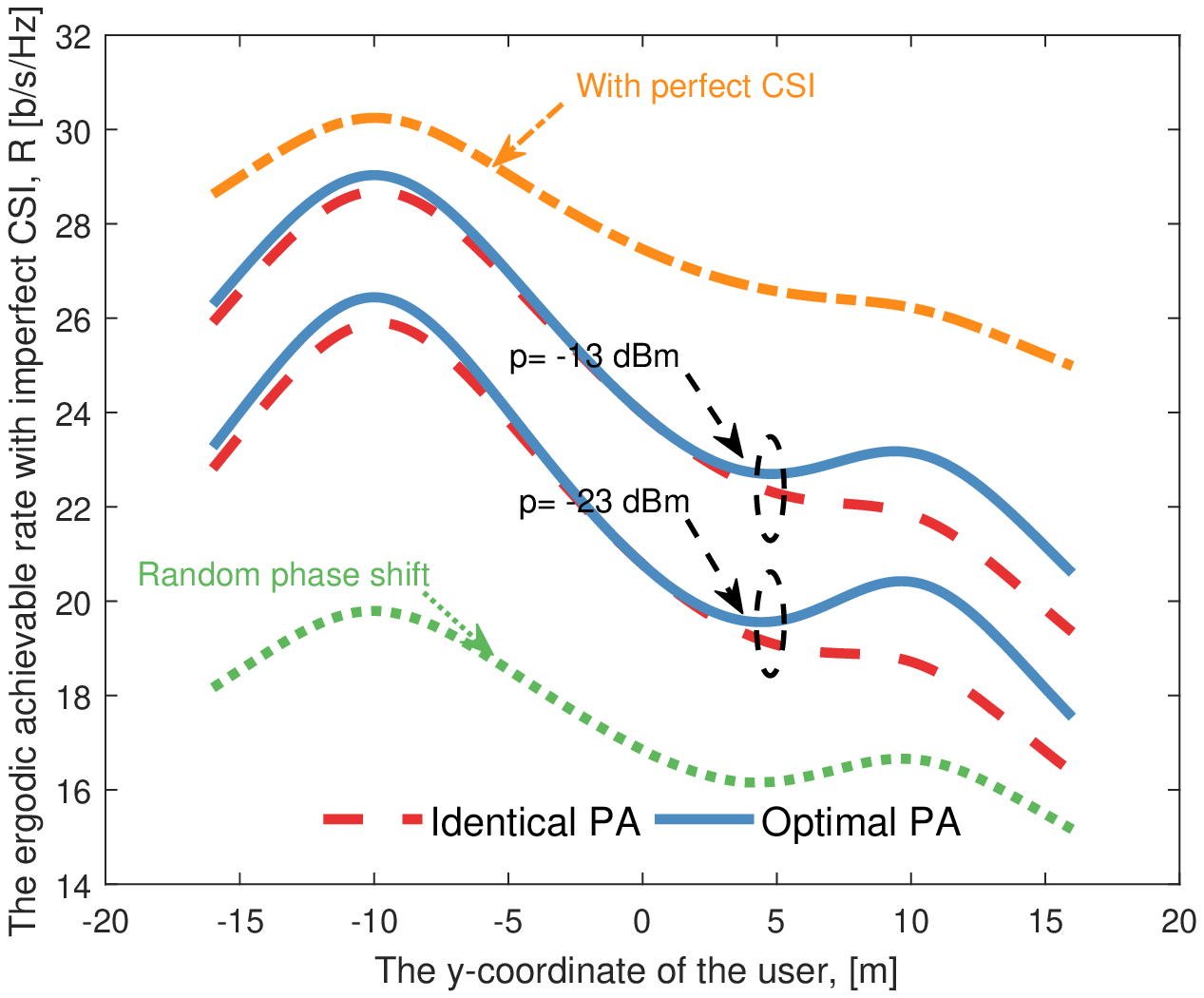}
	\caption{The ergodic achievable rate versus the y-coordinate of the user (asymmetrical arrangement of two IRSs).}
	\label{fig8}
\end{figure}
\begin{figure}[!t]
	\centering
	\includegraphics[width=8.8cm]{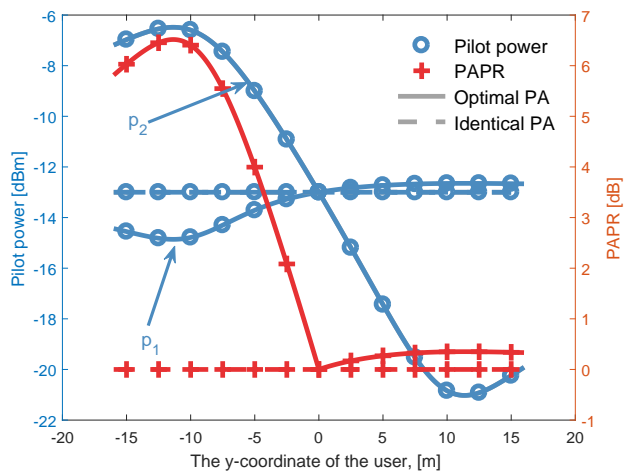}
	\caption{The power and PAPR of pilot signals versus the y-coordinate of the user (asymmetrical arrangement of two IRSs)}
	\label{fig9}
\end{figure} More specifically, when the user moves in the middle of IRS-$1$ and IRS-$2$, the proposed pilot power allocation strategy is equivalent to the identical pilot power allocation strategy, both of which have the PAPR of $0$ dB. However, as the user moves toward the IRS-$1$ or IRS-$2$, the proposed pilot power allocation strategy will incurs the PAPR increase. When the user moves at the location nearest the IRS-$1$/IRS-$2$, the theoretical PAPR can be obtained by ${{2\sqrt {d_{ui1}^{ - {\alpha _{ui}}}} } \mathord{\left/
 {\vphantom {{2\sqrt {d_{ui1}^{ - {\alpha _{ui}}}} } {\left( {\sqrt {d_{ui1}^{ - {\alpha _{ui}}}}  + \sqrt {d_{ui2}^{ - {\alpha _{ui}}}} } \right)}}} \right.
 \kern-\nulldelimiterspace} {\left( {\sqrt {d_{ui1}^{ - {\alpha _{ui}}}}  + \sqrt {d_{ui2}^{ - {\alpha _{ui}}}} } \right)}} \approx 2.4$ dB, which is consistent with the simulation results as shown in Fig. \ref{fig7}. Therefore, to be fair, the rate improvement for the user near the IRS-$1$/IRS-$2$, as shown in Fig. \ref{fig6}, is gained at the compromise of the PAPR increase of about $2.4$ dB.

Furthermore, let us consider the more general cases where IRS-$1$ and IRS-$2$ are equipped with a different number of reflecting elements due to the specific surface area limit. In our simulations, the number of reflecting elements at the IRS-$1$ and IRS-$2$ is set to be ${M_1} = 1000$ and ${M_2} = 100$, respectively. The simulation results are shown in Fig. \ref{fig8}, which is significantly different from that in Fig. \ref{fig6}. As can be seen from \ref{fig8}, the achievable rate of the user near the IRS-$1$ is higher than that near the IRS-$2$, which is due to that more reflecting elements are equipped at the IRS-$1$. Besides, compared to the identical pilot power allocation strategy, the proposed pilot power allocation strategy is more effective in the vicinity of the IRS-$2$, which means that the reflecting elements at IRS-$1$ play a role in improving the capacity of the user near the IRS-$2$. However, for the user in the vicinity of IRS-$1$, the smaller number of reflecting elements at the IRS-$2$ hardly contribute to the rate improvement. In summary, the rate improvement caused by the pilot power allocation is limited in the vicinity of a large IRS.

\begin{figure}[!t]
	\centering
	\includegraphics[width=8.8cm]{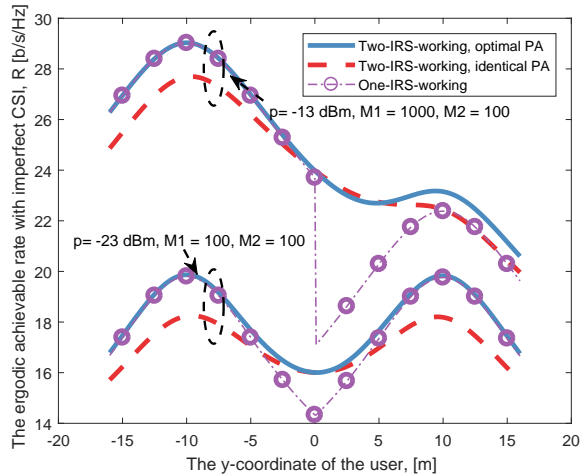}
	\caption{The comparison of the two-IRS-working scheme and the one-IRS-working scheme.}
	\label{fig10}
\end{figure}
Likewise, the specific pilot power and the PAPR comparison for the asymmetric number of reflecting elements are shown in Fig. \ref{fig9}. It can be seen from Fig. \ref{fig9} that the qualitative conclusions that allocating more amount of power to the pilots for estimating the weak reflection channels still holds. Besides, since the number of reflecting elements at IRS-$1$ is much larger than that at IRS-$2$, which makes the pilot power for estimating the reflection channels via IRS-$1$ more similar as the identical pilot power allocation. Furthermore, thanks to such a large IRS-$1$, the user near the IRS-$2$ suffers from slight PAPR increase, e.g., only $0.3$ dB for the user nearest IRS-$2$, but gains significant rate improvement. However, the unfair phenomena appear for the user in the vicinity of the IRS-$1$, where a fairly high PAPR does not boost the rate performance. For example, the PAPR of the pilot signals for the user nearest the IRS-$1$ is as high as $6.5$ dB. Therefore, for the user near the IRS-$1$, the proposed pilot power allocation strategy bring limited performance improvement at the huge increase of PAPR.

In view of that the existence of IRS-$2$ is hardly helpful for the user near the IRS-$1$, we next consider a benchmark scheme, where only a single IRS works when the user moves into its vicinity. The specific simulation parameters are shown in Fig. \ref{fig10}. As shown in Fig. \ref{fig10}, for the case that the same number of reflecting elements are equipped at two IRSs, the proposed optimal pilot power allocation strategy brings limited rate improvement than the single-IRS assisted scheme. However, when the user moves in the middle of two IRSs, the high power loss makes the combination of two IRSs a better option. Furthermore, the asymmetry case with the number of reflecting elements of ${M_1} = 1000,\ {M_2} = 100$ are also shown in Fig. \ref{fig10}. It can be seen from Fig. \ref{fig10} that, when the user is near the IRS-$1$, the single-IRS assisted scheme almost approach the two-IRS assisted scheme with the optimal pilot power allocation. Besides, it is interesting to note that the single-IRS assisted scheme even outperforms the two-IRS assisted scheme with identical pilot power allocation, which implies the worse estimates of the BS--IRS-$2$--user link in the identical pilot power allocation strategy hardly provides rate improvement for the user near the IRS-$1$. In addition, the single-IRS assisted scheme also reduces the pilot overhead from $\left( {{M_1} + {M_2}} \right)$ to ${{M_1}}$ and remains the PAPR of $0$ dB. However, the situation is quite different for the user in the vicinity of the IRS-$2$, where the single-IRS scheme suffers from severe rate loss than the two-IRS scheme. Such a result reflects that a large number of reflecting elements at the IRS-$1$ do help for improving the rate of the user near the IRS-$2$. Therefore, the IRS controller needs to jointly perform the IRS selection and optimal pilot power allocation to gain performance improvement at the minimum cost. Basically, for the user near a large IRS, it is suggested to choose the single-IRS scheme, where only the IRS nearest the user works while other IRSs are switched off. By contrast, when the user moves near the IRS with a small number of reflecting elements, it is better to switch on more IRSs and perform the optimal pilot power allocation strategy to improve the rate performance.

\begin{figure}[!t]
	\centering
	\includegraphics[width=8.8cm]{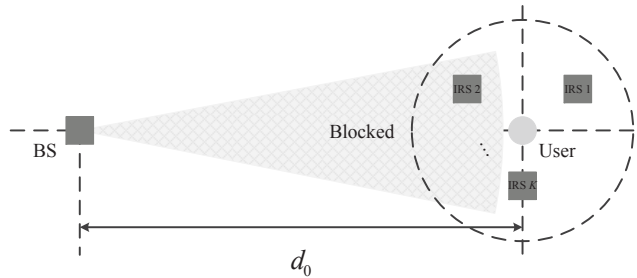}
	\caption{The simulation setup of the multi-IRS assisted communication system (from the top view).}
	\label{fig11}
\end{figure}

Finally, let us consider the general communication system assisted by multiple IRSs, as shown in Fig. \ref{fig11}, where the user is located at $\left( {{d_0},0,0} \right)$ with the horizontal BS-user distance of ${d_0} = 50$ m, while multiple IRSs are randomly located at the vicinity of the user. Specifically, we consider that multiple IRSs are randomly deployed in a circle with the user at the centre (from the top view), the radius is set to be ${d_r} = 20$ m. The height of all IRSs is set to be ${d_h} = 10$ m. In our simulations, we will consider the number of IRSs increases from $K = 1$ to $K = 6$.\begin{figure}[!t]
	\centering
	\includegraphics[width=8.8cm]{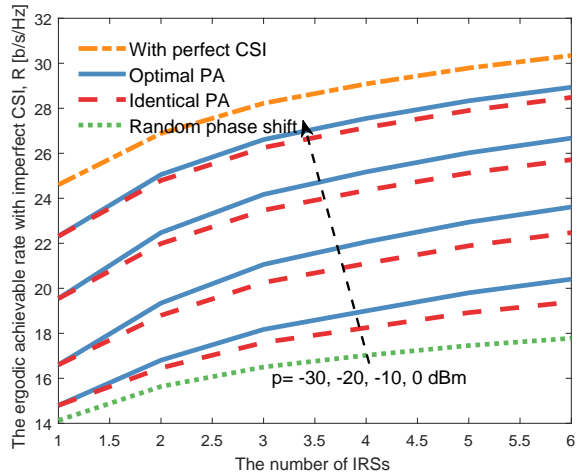}
	\caption{The ergodic achievable rate versus the number of IRSs (with the constraint on the number of reflecting elements at each IRS).}
	\label{fig12}
\end{figure}
\begin{figure}[!t]
	\centering
	\includegraphics[width=8.8cm]{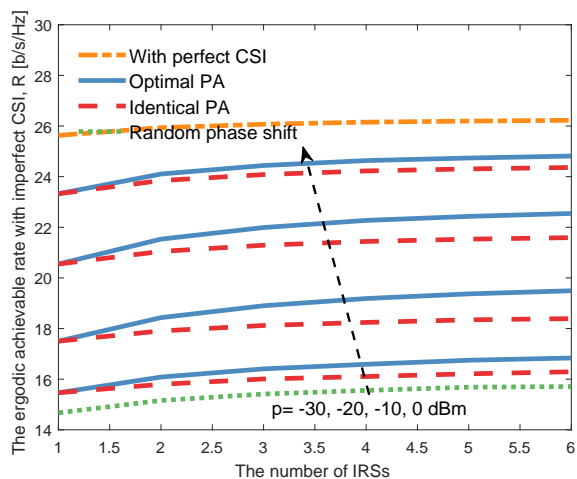}
	\caption{The ergodic achievable rate versus the number of IRSs (with the constraint on the total number of reflecting elements of all IRSs).}
	\label{fig13}
\end{figure} In Fig. \ref{fig12}, we consider that the number of reflecting elements at each IRS is fixed at $1000$, i.e., ${M_1} =  \cdots {M_K} = 1000$. It can be seen from Fig. \ref{fig12} that, with the increase of the number of IRSs, the performance improvement of the optimal pilot power allocation strategy follows the quadratic power law as that with perfect CSI. For example, when doubling the number of IRSs, the rate improvement is about $2$ b/s/Hz, which is readily verified in Fig. \ref{fig12}. By contrast, the identical pilot power allocation strategy suffers from a little rate loss. As the number of IRSs increases, the performance advantages of the proposed pilot power allocation strategy becomes more significant. Besides, the rate performance can be further improved by increasing the average pilot power.

Next, we consider the constraint on the total number of reflecting elements of all IRSs. For the sake of elaboration, we still consider the same number of reflecting elements at each IRS, and the total number of reflecting elements of all IRSs is set to be ${M_{total}} = \sum\nolimits_{k = 1}^K {{M_k}}  = 1440$ to make sure that the total number can be divisible by $K = 1 \sim 6$. Under this constraint, the number of reflecting elements at each IRS is ${M_1} =  \cdots  = {M_K} = {{M_{total}} \mathord{\left/
 {\vphantom {{M_{total}} K}} \right.
 \kern-\nulldelimiterspace} K}$. The simulation results are shown in Fig. \ref{fig13}. It can be seen from Fig. \ref{fig13} that, due to the constraint of the total number of reflecting elements, the rate improvement is limited with the increase of the number of IRSs. Nonetheless, the rate improvement implies that, under the constraint of the total number of reflecting elements, the distributed deployment of IRSs can obtain the improved rate performance as the user moves in the vicinity of them. The reason is that the user can only gain the rate benefit when it moves near one of the IRSs, while the distributed deployment of IRSs can effectively enlarge the effective area of the IRS swarm. In addition, when considering the practical cases, distributed deployment of IRSs can improve the link quality between the IRSs and user. More particularly, when one of the IRS-user links are blocked, other IRSs can also work effectively. Besides, the proposed pilot power allocation remains the most competitive option under all considered simulation settings.

\section{Conclusions}\label{sec7}
In this paper, we proposed the optimal pilot power allocation scheme for multiple-IRS assisted communication systems. Specifically, we first derived the ergodic achievable rate with imperfect CSI, which is dependent on the specific pilot power for estimating different reflection channels. We then formulated the ergodic capacity maximizing problem subject to the constraint of average pilot power and solve the problem by leveraging the method of Lagrange multipliers. The proposed pilot power scheme suggests allocating more amount of power to the pilots used for estimating the weak channel helps to improve the rate performance, which, however, increases the PAPR of pilot signals. The simulation results verified the theoretical ergodic achievable rate and the proposed pilot power allocation strategy. Some useful conclusions are also summarized from our simulations. For example, for the user in the vicinity of a large IRS, it is suggested to turn off other IRSs, while for users near a small IRS, the IRS controller should switch on all IRSs and perform the optimal pilot power allocation. Besides, our simulations have demonstrated that the distributed deployment of IRSs can enlarge the effective area of the IRS swarm, under the constraint of a total number of reflecting elements. Finally, considering that the system model and the continuous phase-shift assumption considered in this paper is relatively primary, the pilot allocation in practical MIMO or multi-user MISO systems with the discrete phase shift model will be left for our future research direction.

\ifCLASSOPTIONcaptionsoff
  \newpage
\fi

\bibliographystyle{IEEEtran}
\bibliography{An}

\end{document}